\documentclass[10pt]{iopart}
\usepackage{graphicx,epsfig,epsf}
\usepackage{bm}
\usepackage{iopams}
\usepackage{comment}
\usepackage{mathbbol}
\usepackage{cite}

\def\ket#1{|#1\rangle}
\def\bra#1{\langle#1|}

\def\xiij{\xi_{ij}}
\def\PRA{\PR \textit{A}}

\makeatletter
\def\bbordermatrix#1{\begingroup \m@th
  \@tempdima 4.75\p@
  \setbox\z@\vbox{%
    \def\cr{\crcr\noalign{\kern2\p@\global\let\cr\endline}}%
    \ialign{$##$\hfil\kern2\p@\kern\@tempdima&\thinspace\hfil$##$\hfil
      &&\quad\hfil$##$\hfil\crcr
      \omit\strut\hfil\crcr\noalign{\kern-\baselineskip}%
      #1\crcr\omit\strut\cr}}%
  \setbox\tw@\vbox{\unvcopy\z@\global\setbox\@ne\lastbox}%
  \setbox\tw@\hbox{\unhbox\@ne\unskip\global\setbox\@ne\lastbox}%
  \setbox\tw@\hbox{$\kern\wd\@ne\kern-\@tempdima\left(\kern-\wd\@ne
    \global\setbox\@ne\vbox{\box\@ne\kern2\p@}%
    \vcenter{\kern-\ht\@ne\unvbox\z@\kern-\baselineskip}\,\right)$}%
  \null\;\vbox{\kern\ht\@ne\box\tw@}\endgroup}
\makeatother

\newcommand{\sza}{\sigma_z^{(i)}}

\newcommand{\spa}{\sigma_+^{(i)}}
\newcommand{\smb}{\sigma_-^{(j)}}
\newcommand{\sma}{\sigma_-^{(i)}}

\begin{document}

\title{Competition between finite-size effects and dipole-dipole interactions in few-atom systems}

\author{Fran\c{c}ois Damanet and John Martin}
\address{Institut de Physique Nucl\'eaire, Atomique et de Spectroscopie, CESAM, Universit\'e de Li\`ege, B\^at.\ B15, B - 4000 Li\`ege, Belgium}

\eads{Francois.damanet@ulg.ac.be}

\begin{abstract}
In this paper, we study the competition between finite-size effects (i.e.\ discernibility of particles) and dipole-dipole interactions in few-atom systems coupled to the electromagnetic field in vacuum. We consider two hallmarks of cooperative effects, superradiance and subradiance, and compute for each the rate of energy radiated by the atoms and the coherence of the atomic state during the time evolution. We adopt a statistical approach in order to extract the typical behavior of the atomic dynamics and average over random atomic distributions in spherical containers with prescribed $k_0R$ with $k_0$ the radiation wavenumber and $R$ the average interatomic distance. Our approach allows us to highlight the tradeoff between finite-size effects and dipole-dipole interactions in superradiance/subradiance. In particular, we show the existence of an optimal value of $k_0R$ for which the superradiant intensity and coherence pulses are the less affected by dephasing effects induced by dipole-dipole interactions and finite-size effects.

\noindent{\it Keywords\/}: superradiance, subradiance, dipole-dipole interactions, decoherence
\end{abstract}

\pacs{42.50.-p,42.50.Nn,03.65.Yz}

\submitto{\jpb}

\maketitle
\ioptwocol

\section{Introduction}

Cooperative processes in atomic systems are of major interest as they occur in a wide range of applications \cite{Bra05, Fic02}. Paradigmatic examples of cooperative processes are supperadiance and subradiance. The former stands for the enhanced - and the latter for the reduced - spontaneous emission of light by excited atoms placed in vacuum. These quantum many-body effects are the subject of intense research for more than sixty years (see e.g.~\cite{Gro82, Men99,Pro06} and references therein). They have recently regained attention in various contexts such as photon localization~\cite{Ack08}, single photon cooperative emission~\cite{Fri10, Li12, Jen16}, non-equilibrium phase transition in dilute thermal gases of Rydberg atoms~\cite{Car13}, cooperative Lamb-shift~\cite{Mei13} or superradiant clock laser~\cite{Mai14}. Superradiance was first predicted by Dicke in his classic paper of 1954~\cite{Dic54}. It is commonly interpreted as a cooperative behaviour, assisted by the electromagnetic field, in which the atoms successively synchronize their dipoles. During this evolution, the  atomic state is restricted to the symmetric subspace of the global Hilbert space and the coherence created during the emission cascade leads to an enhanced spontaneous emission rate.

In the superradiance effect, the perfect synchronization of dipoles is the sole consequence of the indistinguishability of atoms in the sample. As long as it is impossible to tell from which atom a photon is emitted, the various de-excitations paths interfere together. Constructive interference gives rise to superradiance whereas destructive interference leads to subradiance. Subradiant states are of particular interest for atomic implementations of qubit systems as they can be immune against decoherence due to spontaneous emission \cite{Lid14}. 

For atoms to be indistinguishable, two requirements must be met: i) the interatomic distance should be much smaller than the wavelength $\lambda$ of the emitted radiation, and ii) each atom should experience the same dipole-dipole shift due to the surrounding atoms. As soon as one of these requirements is not met, superradiance/subradiance will be altered \cite{Cof78,Fri74, Fre86, Fre87, Fen13, Ric90, Fri72, Kar07, Bie12}. In both cases, a dephasing between the atomic dipoles will occur which has the effect of coupling the global state vector to states of lower symmetry and reducing the coherence.

The effects of the finite size of the atomic sample, i.e.\ of the breakdown of condition i), have been studied by many authors. They can be accounted for by introducing the cooperativity parameter, $\mathcal{C} = \rho \lambda^3/4\pi^2$ with $\rho$ the number density, quantifying the reduction in emission rates~\cite{Wan07,Lin12}. For closely packed atoms, $\mathcal{C}\gg 1$ and superradiance is pronounced, while it is suppressed for $\mathcal{C}\ll 1$. 

The effects of dipole-dipole interactions, i.e.\ of the breakdown of condition ii), have also been studied in great detail \cite{Lid14,Yav14}. They are significant when the atomic distribution is no longer invariant under permutation of the particles. Hence, they are present as soon as the number of atoms exceeds two. The description of these effects involves a huge amount of degree of freedom as compared to the non-interacting case and it is therefore a hard task to solve the equations describing the atomic dynamics. In the literature, several methods have been proposed, such as effective two-atom master equation~\cite{Wan07, Lin12b, Fle99}, quantum trajectory approach \cite{Cle02, Cle03, Cle04, Cle08} and optimally convergent quantum jump expansion \cite{Luc13, Luc14}. Remarkably, some analytical solutions exist \cite{Fre86,Fre87,Fen13}, however their expressions are often quite involved except in few particular cases, which makes it difficult to extract the main features of these effects.

In this paper, we study the combined effects of the breakdown of conditions i) and ii) on superradiance and subradiance. This is particularly interesting as these conditions are generally incompatible: small interatomic distances lead to small finite-size effects but to large dipole-dipole interactions. We provide quantitative results about the impact of dipolar interactions on the dynamics of a finite-size atomic sample of $3, 4$ and $5$ atoms randomly distributed in space. By considering a small number of atoms, it is still tractable to solve numerically the full master equation to describe the complex interplay between dipole-dipole interactions and finite-size effects on the dissipative atomic dynamics. This allows us to analyse the time evolution of the radiated energy rate and of a proper measure of coherence of the atomic system for two different initial states: a fully-excited state and a decoherence-free state (in the absence of dipolar interactions). As most experimental works have focused so far on superradiance in samples containing a large number of atoms, the exact study of few-atom systems provides complementary information about their dynamics that might reveal useful for future experiments that could be realized with current technology, e.g.\ with trapped Rydberg atoms~\cite{Bro16}.


The paper is organized as follows. In Section II, we describe the system under investigation and the master equation governing its dynamics. In Section III, we explain our method to study the combined effects of finite-size and dipole-dipole interactions on cooperative processes. Section IV is dedicated to our results for the two different situations considered above: superradiance and subradiance. The Appendix A contains the technical details about the method we used to solve the full master equation. We present our conclusion in Section V.

\section{System and master equation}

\begin{figure}[h!]
  \begin{center}
	\includegraphics[scale=0.8]{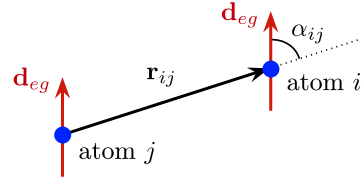}
    \caption{Angle $\alpha_{ij}$ between the relative position $\mathbf{r}_{ij} = \mathbf{r}_i - \mathbf{r}_j$ of atoms $i$ and $j$ and the atomic dipole moment $\mathbf{d}_{eg}$.}
    \label{fig:alphaij}
   \end{center}
\end{figure}

We consider a system of $N$ identical two-level atoms at fixed positions $\mathbf{r}_i$ ($i = 1, \dots, N$) coupled to the quantized electromagnetic field at zero temperature (vacuum). We denote by $\hbar \omega_{0}$ ($\omega_0 = 2\pi c/\lambda_{0} = k_0 c$) the energy difference between the excited state $\ket{e_i}$ and the ground state $\ket{g_i}$ of atom $i$. In the dipole approximation, atoms are modelled as point-dipoles with electric dipole moment $\mathbf{d}^{(i)}_{eg}=\bra{e_i}\mathbf{d}\ket{g_i}$. We assume that the atomic sample is polarized, e.g.\ by an external field, so that all atoms have the same dipole moment, $\mathbf{d}^{(i)}_{eg}=\mathbf{d}_{eg}\;\forall\,i$. We consider that the atomic levels are coupled by $\pi$-polarized light, so that $\mathbf{d}_{eg}$ can be taken real. The internal dynamics of the atoms is governed by a Markovian master equation for the density matrix $\rho(t)$~\cite{Lin76}. In the interaction picture with respect to the system Hamiltonian $H_S= (\hbar \omega_0/2) \sum_{i=1}^{N} \sza$ with $\sza = \ket{e_i}\bra{e_i}-\ket{g_i}\bra{g_i}$, it reads~\cite{Aga74} 
\begin{equation}\label{me}
\frac{d\rho(t)}{dt} = - \frac{i}{\hbar}\big[ H_{\mathrm{dd}} , \rho(t) \big] + \mathcal{D}\left(\rho(t)\right) \equiv \mathcal{L}\left(\rho(t)\right)
\label{eq:me}
\end{equation}
with
\begin{equation}
H_{\mathrm{dd}} = \sum_{i \neq j}^{N} \hbar f_{ij}\, \spa\smb,
\label{LSH}
\end{equation}
and
\begin{equation}
\label{dissipator}
\mathcal{D}\left(\rho\right) = \sum_{i,j = 1}^{N} \gamma_{ij} \left( \smb \rho \spa - \frac{1}{2} \left\{  \spa\smb, \rho \right\} \right),
\end{equation}
where $\spa = \ket{e_i}\bra{g_i}$ and $\sma = \ket{g_i}\bra{e_i}$ are the raising and lowering operators for atom $i$. The coefficients $f_{ij}$ and $\gamma_{ij}$ entering the master equation are respectively given by~\cite{Ste64,Leh70,Aga74} 
\begin{equation}
\label{eq:fij}
\eqalign{
f_{ij} = \frac{3\gamma_0 }{4} \Bigg[ & (1-3 \cos^2 \alpha_{ij}) \left( \frac{\sin\xiij}{\xiij^2} + \frac{\cos\xiij}{\xiij^3}\right) \cr & \quad - (1 - \cos^2 \alpha_{ij}) \frac{\cos\xiij}{\xiij} \Bigg]},
\end{equation}
and
\begin{equation}
\label{eq:gammaij}
\eqalign{
\gamma_{ij} = \frac{3 \gamma_0}{2} \Bigg[ & (1-3 \cos^2 \alpha_{ij})  \left( \frac{\cos \xiij}{\xiij^2} - \frac{\sin \xiij}{\xiij^3}\right) \cr & \quad + (1 - \cos^2 \alpha_{ij}) \frac{\sin\xiij}{\xiij} \Bigg]}
\end{equation}
with $\xi_{ij}=k_0r_{ij}$, $\gamma_0=(\omega_0^3d_{eg}^2)/(3\pi\hbar\epsilon_0c^3)$ the single-atom spontaneous emission rate and $\alpha_{ij}$ the angle between the relative position $\mathbf{r}_{ij} = \mathbf{r}_i - \mathbf{r}_j$ of atoms $i$ and $j$ and the atomic dipole moment $\mathbf{d}_{eg}$ (see Fig.~\ref{fig:alphaij}). Note that $|\gamma_{ij}|\leqslant\gamma_{ii}=\gamma_0\;\forall\,i,j$.

The Hamiltonian $H_{\mathrm{dd}}$ (Eq.~(\ref{LSH})) describes the conservative dipolar interactions between neutral atoms, that can be interpreted as virtual transverse photon exchanges between excited and ground state atoms \cite{Ste64,Gro82}. The dissipator $\mathcal{D}\left(\rho\right)$ (Eq.~(\ref{dissipator})) accounts for dissipation, i.e.\ photon emission. 

The internal dynamics of the atoms depend both on their relative positions $\mathbf{r}_{ij}$ and on the atomic transition wavelength $\lambda_0$ through the parameters $\xi_{ij}=2\pi\,r_{ij}/\lambda_0$ and $\alpha_{ij}$ on which the coefficients $\gamma_{ij}$ and $f_{ij}$ depend (see Eqs.~(\ref{eq:fij}) and (\ref{eq:gammaij})). In particular, the rate of energy released by the atoms is known to depend drastically on the dimensionless parameter $\xi_{ij}$~\cite{Gro82}. Two limiting regimes appear when this ratio is large or small compared to unity. For the sake of clarity and to introduce notations, we briefly outline the main features associated to these two regimes.

For distant atoms ($\xiij \gg 1$), one has
\begin{equation}
f_{ij} \approx 0,  \quad \gamma_{ij} \approx \gamma_0\, \delta_{ij} \qquad \forall\,i,j.
\end{equation}
In this regime, dipolar interactions are negligible and the master equation (\ref{eq:me}) describes $N$ independent emitters. This is reflected by the fact that the positive semidefinite matrix $\gamma$ with entries $\gamma_{ij}$ has a $N$-fold degenerate eigenvalue equal to $\gamma_0$, the decay rate of an isolated atom. Superradiance and subradiance are completely suppressed in this regime.

For spatially close atoms ($\xiij \ll 1$), one has
\begin{equation}
f_{ij} \approx  \frac{3\gamma_0}{4} \frac{(1-3 \cos^2 \alpha_{ij})}{\xiij^3} ,  \quad
\gamma_{ij} \approx \gamma_0 \qquad \forall\,i,j.
\label{eq:fijcloseatoms}
\end{equation}
In this regime, cooperative effects play a prominent role and collective spontaneous emission processes, such as superradiance or subradiance, are observable. The Hamiltonian $H_{\mathrm{dd}}$ with $f_{ij}$ given by Eq.~(\ref{eq:fijcloseatoms}) accounts for \emph{static} dipole-dipole interactions, with their characteristic $1/r_{ij}^3$ dependence. When $\gamma_{ij} = \gamma_0\;\forall\,i,j$, the dissipator takes the form
\begin{equation}
\mathcal{D}\left(\rho \right) = \gamma_0 \left( J_- \rho J_+ - \frac{1}{2} \Big\{  J_+J_-, \rho \Big\} \right)
\end{equation}
where $J_{\pm} =\sum_{i=1}^{N} \sigma_{\pm}^{(i)}$ are collective spin raising and lowering operators. This particular form of the dissipator, in terms of collective spin operators, highlights the indistinguishability of atoms regarding dissipation processes and preserves the symmetry of the atomic state. From a physical point of view, it is impossible to track down the atom that has emitted a photon when $\xiij \ll 1$, because then the wavelength of the radiation is much larger than the size of the atomic sample. The atomic state evolves in the symmetric subspace of the global Hilbert space, spanned by the Dicke states $\ket{JM}$ of maximal cooperation number $J = N/2$. Dicke states are defined as simultaneous eigenstates of $J^2 = J_x^2 + J_y^2 + J_z^2$ and $J_z$ where $J_m = \frac{1}{2}\sum_{i=1}^{N} \sigma_{m}^{(i)}\;\; (m = x,y,z)$ are the collective spin operators. On the one hand, when the atoms are initially in the fully excited state $|N/2\,N/2\rangle=|e\ldots e\rangle$, the time evolution is a cascade down the $\ket{JM}$ ladder, which is commonly interpreted as a consequence of the phase-synchronization of the atomic dipole moments~\cite{Gro82}. The enhanced rate at which this evolution occurs can be related to constructive interferences between multiple emission paths~\cite{Wie11}. On the other hand, when the atoms are initially in a subradiant state $|J\,-J\rangle$, destructive interferences lead to vanishing decay rates (i.e.\ dark states)~\cite{Bei00}.

However, apart from these two limiting regimes, dephasing between atomic dipoles occurs during the dynamics. Dipole-dipole interactions lead to excitation trapping in the system, thereby contributing to a decrease of the photon emission rate. The decay rates, given by the real part of the eigenvalues of the Liouvillian $\mathcal{L}$, depend on the particular atomic arrangement through the dipole-dipole shifts $f_{ij}$ and dissipation rates $\gamma_{ij}$. 

In the following, we shall refer to the \emph{pure superradiant regime} as the regime in which
\begin{equation}\label{puresup}
f_{ij} \approx  f_0 , \quad
\gamma_{ij} \approx \gamma_0 \qquad \forall\,i,j.
\end{equation}

\section{Method}
In the general case of random atomic distributions, dipole-dipole interactions couple symmetric states (in particular the fully excited state) to states with lower symmetry. 

The number of available states increases exponentially with the number of atoms. When analytical solutions exist, (see e.g.~\cite{Cof78,Fre86, Fre87,Fre04,Fen13}), they generally exhibit complicated expressions and the main features of the dynamics are difficult to extract from them. For this reason, we choose to adopt a statistical approach in order to extract the typical behavior of the atomic dynamics. Our numerical procedure is as follows. We generate random atomic distributions in spherical containers with an average interatomic distance $R$. For each distribution, we compute the radiated energy rate (\ref{eq:int}) and the coherence (\ref{eq:coh}) from the full solution of the master equation (\ref{eq:me}). Finally, we compute average values over all distributions (typically a few thousands, see Appendix for further details) of the radiated energy rate and the coherence defined below. In our simulations, all random atomic distributions are characterized by the same magnitude of the key parameter $k_0R$. The procedure is then repeated for different values of $k_0R$. Since dipole-dipole interactions do not alter the atomic dynamics when there are only $2$ atoms, we focus on systems made of $3$, $4$ and $5$ atoms in the following.

\section{Results and discussion}

In this section, we present our results on the influence of dipole-dipole interactions and finite-size effects on cooperative processes. In order to characterize the dynamics of the atomic system, we compute the normalized radiated energy rate
\begin{equation}
I(t) = -\frac{2}{\hbar \omega_0}\frac{d}{dt} \Tr  [H_{S}\rho(t)]
\label{eq:int}
\end{equation}
and the $l_1$\emph{-norm  of coherence} $C_{l_1} (t)$ defined as
\begin{equation}
C_{l_1} (t) = \sum_{m,n \atop m \neq n} | \rho_{mn}(t) |
\label{eq:coh}
\end{equation}
where $\rho_{mn}(t)$ $(m,n = 1, \dots , 2^N)$ are the density matrix elements in the basis formed by all combinations of tensor products of individual atomic states $\ket{g_i}$ and $\ket{e_i}$ ($i=1,\ldots,N$). The radiated energy rate provides information on how fast the energy is released in the environment and its behavior is an indicator of the presence of superradiance commonly used in the literatture. With the $l_1$\emph{-norm  of coherence}~\cite{Bau13}, one can quantify the build-up and the fading of coherence in the atomic system.

In the regime of cooperative emission ($k_0 r_{ij} \lesssim 1$), the system evolution depends drastically on the atomic arrangement through the dipole-dipole shifts $f_{ij}$. It is only when the atoms are far apart ($k_0 r_{ij} \gg 1$) that dipole-dipole interactions become negligible and the master equation (\ref{me}) describes $N$ independent emitters. In this regime, the fully excited initial state
\begin{equation}
\rho(t_0 = 0) = \ket{e_1\dots e_N}\bra{e_1 \dots e_N}
\label{eq:fe}
\end{equation}
leads to a radiated energy rate that decreases exponentially with time and a coherence that remains zero at any time, i.e.\
\begin{equation}
I(t) = N\gamma_0\, e^{-\gamma_0 t}, \qquad C_{l_1}(t) = 0.
\label{eq:ispont}
\end{equation}
In this case, the only stationary state is the ground state $\ket{g_1\dots g_N}$.

In the next two subsections, we consider two different initial states: a fully-excited state leading to superradiance, and a decoherence-free state with respect to spontaneous emission leading to subradiance.

\subsection{Fully-excited state : superradiance}

\subsubsection{Identical dipole-dipole shifts $f_{ij}$}

When all dipole-dipole shifts $f_{ij}$ are identical, the atomic state evolves up to a global phase factor as in the absence of dipole-dipole shifts. It is restricted to the symmetric subspace spanned by the $N+1$ symmetric Dicke states $\ket{JM}$ with $J = N/2$ and $M = -J, \dots, J$. In this case, dipole-dipole interactions have no impact on superradiance nor on subradiance~\cite{Li12, Dic54, Gro82, Fri74, Cof78, Fen13}. 

For $N = 3$ atoms, the radiated energy rate (\ref{eq:int}) and the coherence (\ref{eq:coh}) can be calculated analytically for an initial fully excited state. They read
\begin{equation}
I(t) = 3 \gamma_0\, e^{-4 \gamma_0 t} \left[ 8 + e^{\gamma_0 t} (12 \gamma_0 t-7) \right],
\label{eq:isuperradianceN3}
\end{equation} 
\begin{equation}
C_{l_{1}}(t) = 3\, e^{-4 \gamma_0 t} \left[ 6 + e^{ \gamma_0 t} (8 \gamma_0 t-6) \right].
\label{eq:csuperradianceN3}
\end{equation}
The intensity is $3\gamma_0$ at $t=0$, increases with time and reaches a maximum $I_\mathrm{max}\approx 3.225 \gamma_0$ at $t\approx 0.157/\gamma_0$ before decreasing to $0$. 
As for the coherence, it is zero at $t=0$, increases with time and reaches a maximum $C_{l_{1},\mathrm{max}}\approx 1.109$ at $t\approx 0.438/\gamma_0 $ before decreasing to $0$. The build-up and fading of coherence is thus a characteristic trait of superradiance. This behavior is illustrated in Fig.~2 (dotted curves).

For larger number of atoms, it is still possible to find analytical expressions for $I(t)$ and $C_{l_{1}}(t)$, although more involved.
For $N = 4$, we find
\begin{equation}
I(t) = \gamma_0\, e^{-6 \gamma_0 t} \left[72 \gamma_0 t+4 e^{2\gamma_0 t} (36 \gamma_0 t-23)+96\right],
\label{eq:isuperradianceN4}
\end{equation}
\begin{equation}
C_{l_{1}}(t) = 12 e^{-6 \gamma_0 t} \left[4 \gamma_0 t+e^{2 \gamma_0 t} (9 \gamma_0 t-6)+6\right]
\label{eq:csuperradianceN4}
\end{equation}
and for $N = 5$,
\begin{equation}
\eqalign{
I(t) = \gamma_0\, \frac{5}{3} e^{-9 \gamma_0 t} \Big[ & 16 e^{\gamma_0 t} (24 \gamma_0 t-1) \cr 
& \;\; +e^{4 \gamma_0 t} (240 \gamma_0 t-143)+162\Big],}
\label{eq:isuperradianceN5}
\end{equation}
\vspace*{-10pt}
\begin{equation}
C_{l_{1}}(t) = \frac{20}{3} e^{-8 \gamma_0 t} \left[5 (6\gamma_0 t+5)+e^{3\gamma_0 t} (48 \gamma_0 t-25)\right].
\label{eq:csuperradianceN5}
\end{equation}
The maxima of intensity are $I_\mathrm{max}\approx 4.857 \gamma_0$ for $N=4$ and $I_\mathrm{max}\approx 6.879 \gamma_0$ for $N=5$ and occur respectively at $t\approx 0.214/\gamma_0$ and $t\approx 0.233/\gamma_0$. As for the coherence, its maxima are $2.460$ at $t\approx 0.390/\gamma_0$ for $N=4$ and $4.892$ at $t\approx 0.352/\gamma_0$ for $N=5$.

\subsubsection{No identical dipole-dipole shifts $f_{ij}$}

\begin{figure}[h!]
\begin{center}
\includegraphics[scale=0.6]{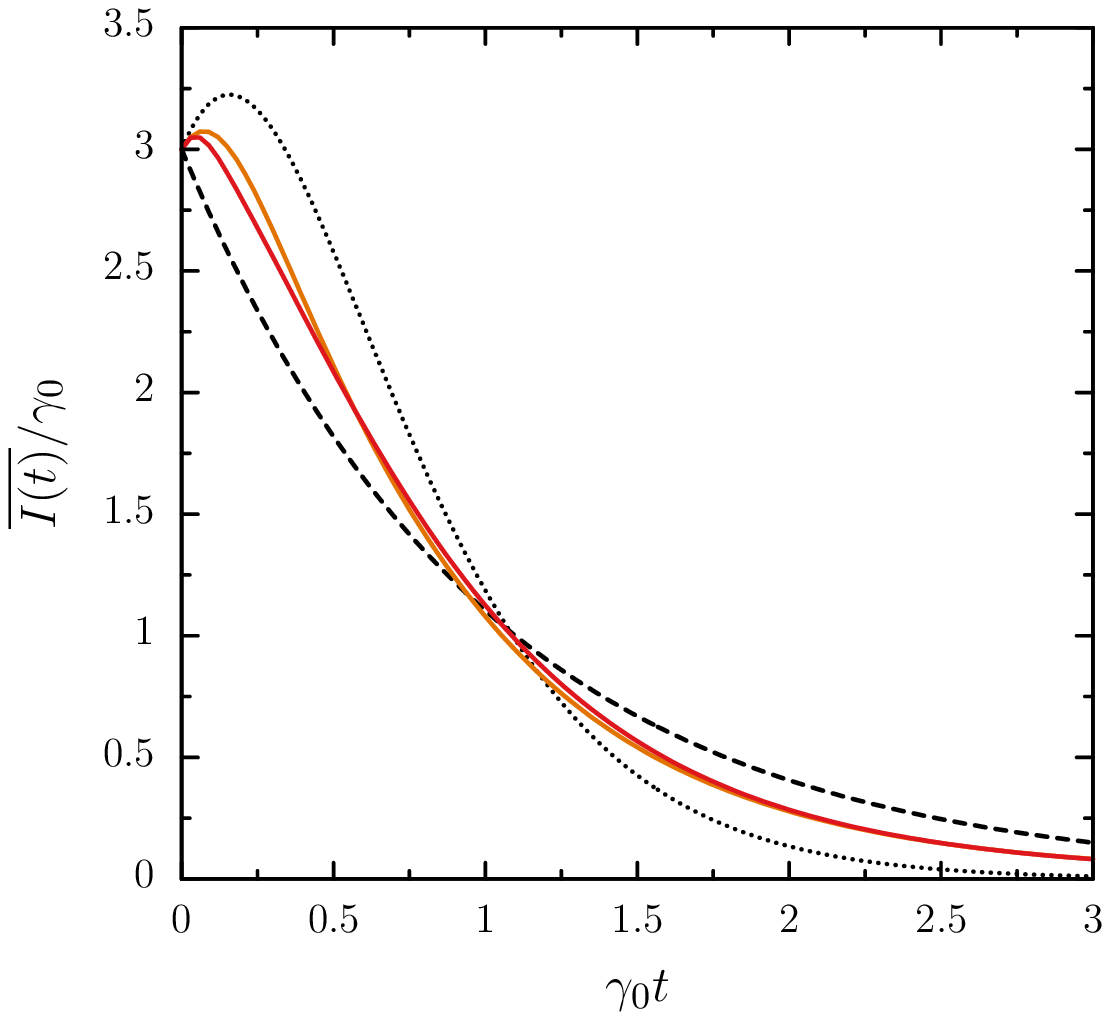}\\[6pt]
\includegraphics[scale=0.6]{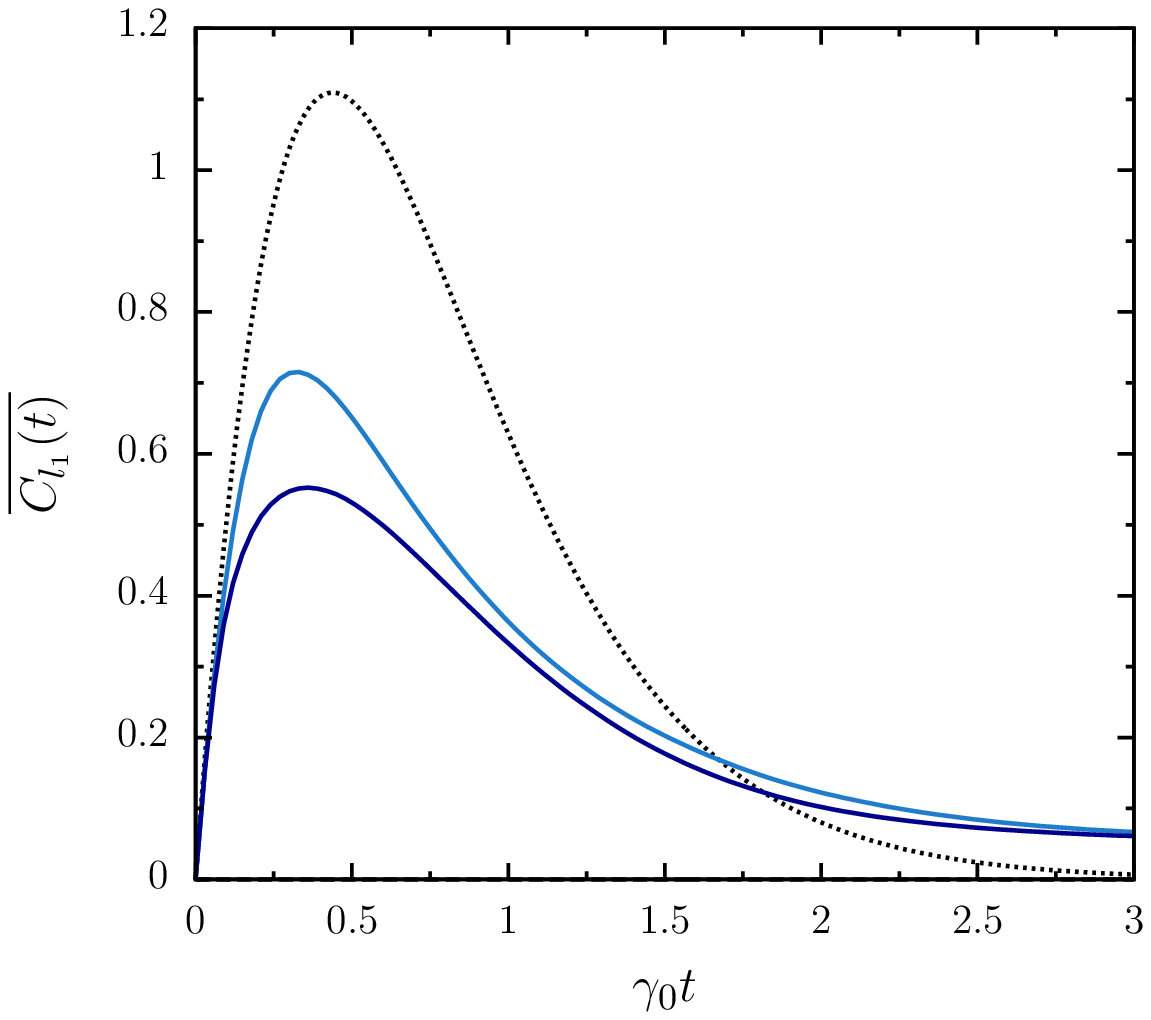}
\caption{Statistical average of the radiated energy rate ($\overline{I(t)}$, top) and the coherence ($\overline{C_{l_1}(t)}$, bottom) for $3$ atoms initially in the fully excited state $|e_1,e_2,e_3\rangle$. Average values are taken over $5000$ random atomic distributions (see method in the Appendix). The dotted and dashed curves show respectively pure superradiance (Eqs.~(\ref{eq:isuperradianceN3}) and (\ref{eq:csuperradianceN3})) and independent spontaneous emission (Eq.~(\ref{eq:ispont})). The light (dark) solid curve shows $\overline{I(t)}$ (top) and $\overline{C_{l_1}(t)}$ (bottom) for $k_0R=0.651$ ($k_0R = 0.466$). The radiated energy rate and the coherence are both altered by dephasing of the atomic dipoles caused by asymmetric dipolar interactions and by finitie-size effects.}
\label{fig:intcoh_behavior}
\end{center}
\end{figure}

\begin{figure}
\begin{center}
\includegraphics[scale=0.6]{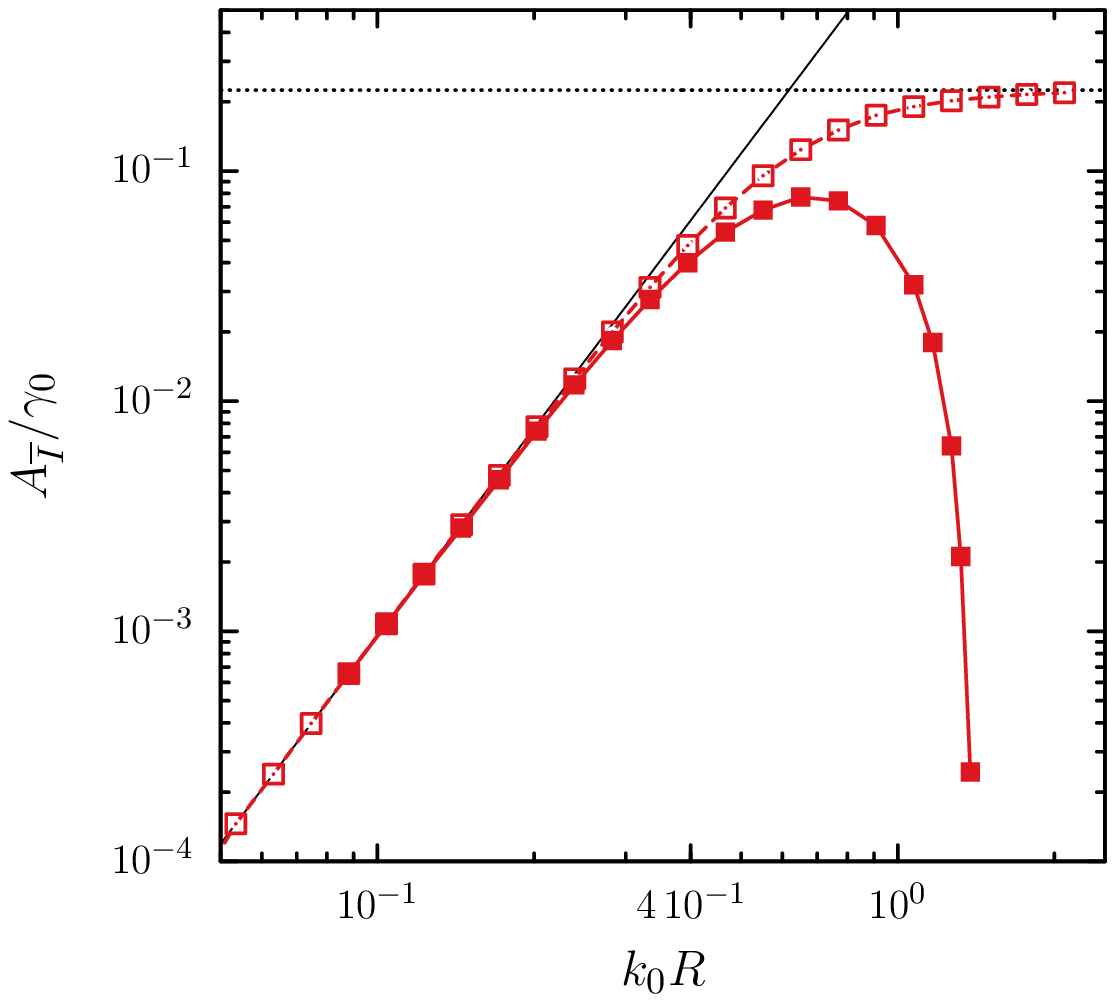}\\[10pt]
\includegraphics[scale=0.6]{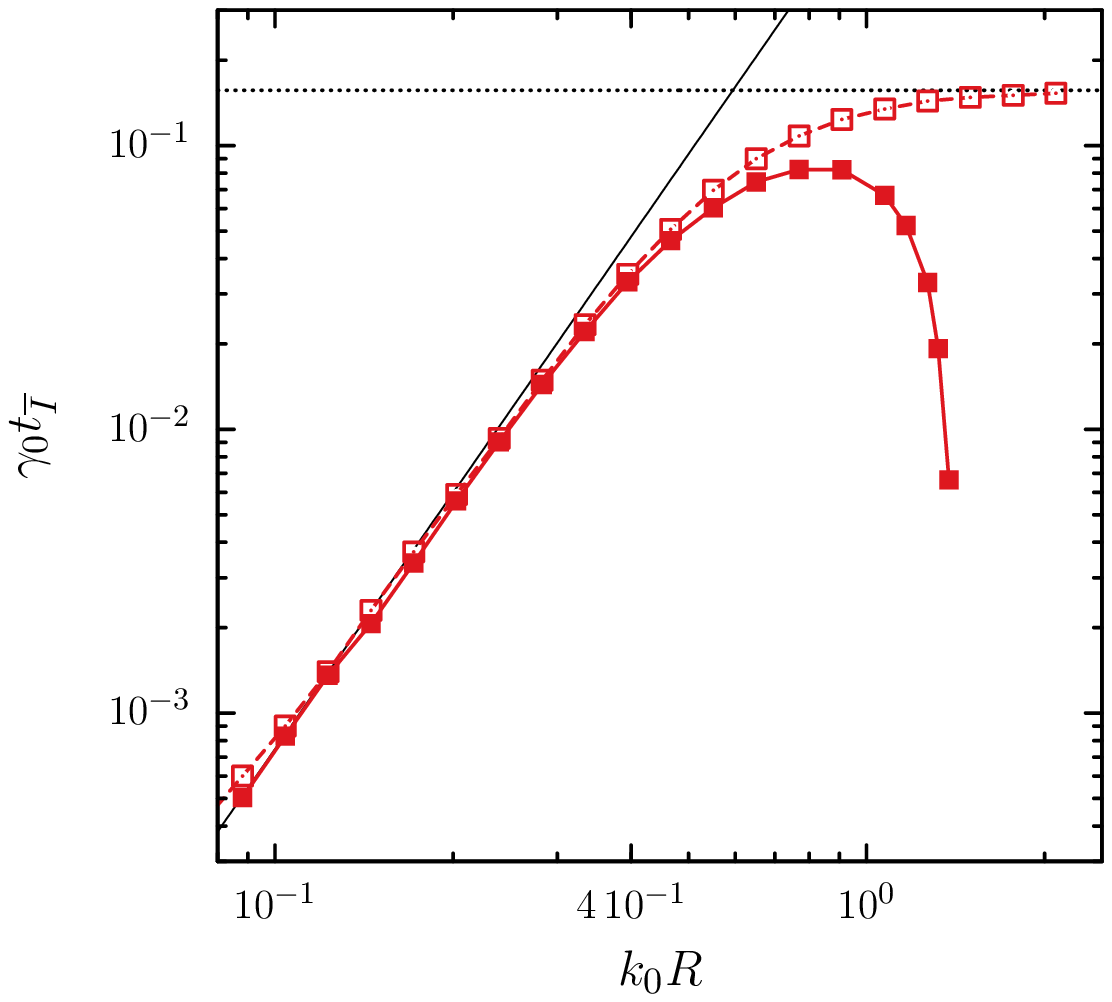}
\caption{Relative maximum $A_{\overline{I}}$ of the average radiated energy rate (top) and time $t_{\overline{I}}$ of the maximum of $A_{\overline{I}}(t)$ (bottom) as a function of $k_0R$ for $N = 3$. Full squares are the results obtained from the solution of the master equation (\ref{eq:me}) with the exact coefficients $\gamma_{ij}$ and $f_{ij}$ given in Eqs.~(\ref{eq:fij})-(\ref{eq:gammaij}). Empty squares, plotted for comparison, are the results obtained with the approximated coefficients (\ref{eq:fijcloseatoms}) valid in the limit of spatially close atoms (i.e.\ for $k_0R\ll 1$). As expected, the full squares collapse with the empty squares for $k_0R\ll 1$. However, the exact results (full squares) display a qualitatively different behavior for moderate and large values of $k_0R$. In particular, $A_{\overline{I}}$ displays a maximum at $k_0R\approx 0.67$. The dotted line shows the value predicted for pure superradiance, i.e.\ \emph{without} dipole-dipole interactions (Eq.~(\ref{puresup})).}
\label{fig:fe_res1}
\end{center}
\end{figure}

Figure~\ref{fig:intcoh_behavior} shows the averages over many random realizations of the radiated energy rate and coherence, $\overline{I(t)}$ and $\overline{C_{l_1}(t)}$, with respect to time for different values of $k_0R$. The radiated energy rate takes the form of a superradiant pulse, although with a reduced amplitude as compared to pure superradiance. The coherence also displays a pulse-like behavior.
In order to characterize quantitatively the reduction of radiated energy rate and coherence caused by dipole-dipole interactions and finite-size effects, we compute the relative maxima of $\overline{I(t)}$ and $\overline{C_{l_1}(t)}$ (hereafter simply called maxima) defined by
\begin{eqnarray}
A_{\overline{I}} = \max_{t \geqslant 0}{\left(\overline{I(t)}\right)} -  I_0 \;\equiv\; \overline{I(t_{\overline{I}})}  -  I_0 \label{Abar}\\
A_{\overline{C}} = \max_{t \geqslant 0}{\left(\overline{C_{l_1}(t)}\right)} - C_{l_1,0} \;\equiv\; \overline{C_{l_1}(t_{\overline{C}})} - C_{l_1,0} \label{Cbar}
\end{eqnarray}
where $I_0 = N \gamma_0$ and $ C_{l_1,0}=0$ (these are the initial values corresponding to the pure superradiant regime). Figure~\ref{fig:fe_res1} (top) shows the change of intensity maximum $A_{\overline{I}}$ with $k_0R$ in the case of three atoms. For short interatomic distances ($k_0R \lesssim 1$), dipole-dipole interactions dominate ($f_{ij}\gtrsim \gamma_0$) and the superradiant pulse is not very pronounced. For large interatomic distances ($k_0R \gtrsim 1$), dipole-dipole interactions are negligible but atoms become distinguishable (large finite-size effects leading to $\gamma_{ij}\lesssim \gamma_0$ for $i\ne j$) and the superradiant pulse is not very pronounced either. In the intermediate regime, there is a trade-off between dipole-dipole interactions and finite-size effects giving rise to a maximum of average radiated intensity rate $A_{\overline{I},\mathrm{max}}\approx 0.078\,\gamma_0$ located at $k_0R \approx 0.67$ (to be compared to $0.225\,\gamma_0$ for pure superradiance). No such maximum is predicted on the basis of the master equation (\ref{eq:me}) with approximated coefficients (\ref{eq:fijcloseatoms}) as is shown by the empty squares in Fig.~\ref{fig:fe_res1} (top) (see~also \cite{Cof78, Gro82}). Moreover, our numerics show that there exists a threshold value $k_0R\approx 1.3$ above which no superradiance occurs (i.e.\ $\overline{I(t)}<I_0\;\forall\,t$). Above the threshold, dephasing processes and finite size effects completely destroy the superradiant pulse. The time after which $\overline{I(t)}$ is maximum is shown in Fig.~\ref{fig:fe_res1} (bottom). It also displays a maximum $t_{\overline{I},\mathrm{max}}\approx 0.083/\gamma_0$ located at $k_0R\approx 0.83$ (to be compared to $0.157/\gamma_0$ for pure superradiance). At small $k_0R$, both the relative superradiant pulse maximum $A_{\overline{I}}$ and the delay time $t_{\overline{I}}$ fall off as $(k_0R)^3$, in agreement with Ref.~\cite{Cof78}. In this regime, the dynamics is dominated by dipole-dipole interactions and the typical decay time is increased by a factor $\sim 1/(k_0 R)^{3}$, while the population transfer between states of different excitation numbers is slowed down and the radiated energy rate is reduced by a factor $\sim (k_0R)^3$. Our results confirm quantitatively this effect.

\begin{figure}[h!]
\begin{center}
\includegraphics[scale=0.6]{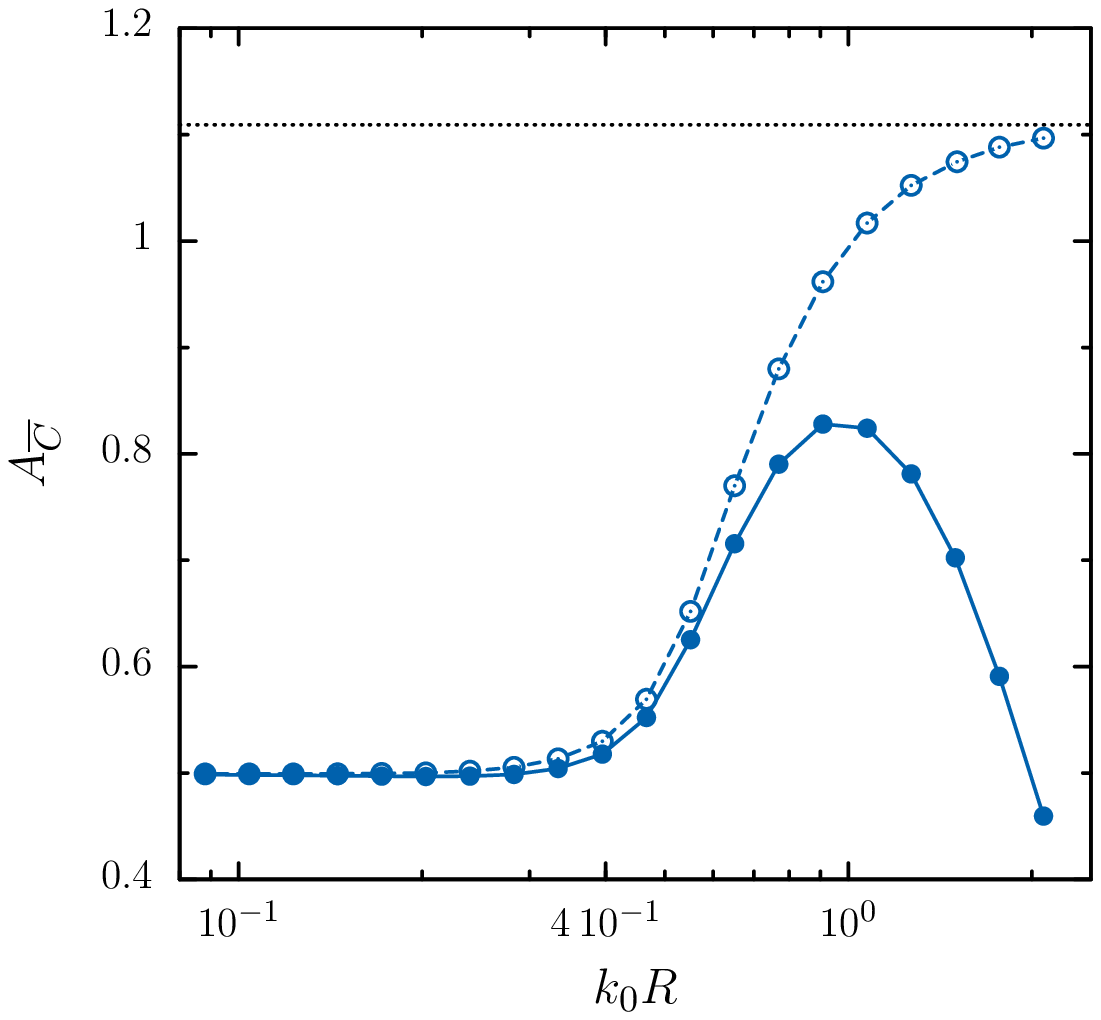}\\[10pt]
\includegraphics[scale=0.6]{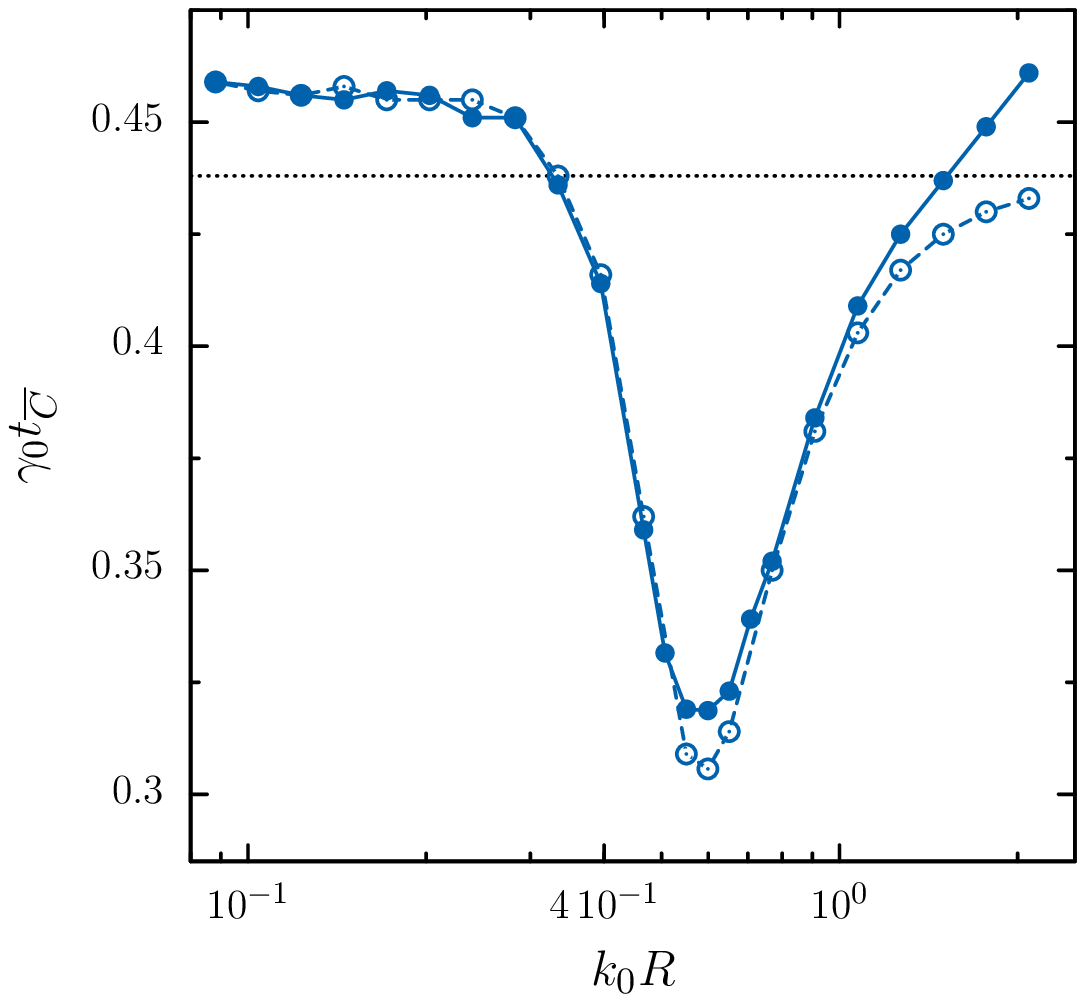}
\caption{Relative maximum $A_{\overline{C}}$ of the average coherence (top) and time $t_{\overline{C}}$ of the maximum of $A_{\overline{C}}(t)$ (bottom) as a function of $k_0R$ for $N = 3$. Full (empty) circles are data obtained from the solutions of the master equation (\ref{eq:me}) with exact (approximated) coefficients $\gamma_{ij}$ and $f_{ij}$ given in Eqs.~(\ref{eq:gammaij}) and (\ref{eq:fij}) (Eq.~(\ref{eq:fijcloseatoms})). The dotted lines show the values predicted for pure superradiance (Eq.~(\ref{puresup})).} 
\label{fig:fe_res3}
\end{center}
\end{figure}

Figure~\ref{fig:fe_res3} (top) shows the change of coherence maximum $A_{\overline{C}}$ with $k_0R$. As for $A_{\overline{I}}$, the average coherence displays a maximum $A_{\overline{C},\mathrm{max}}\approx 0.83$ which is now located at $k_0R \approx 1.00$. This reveals the connection between coherence and enhanced emission rate in the presence of dipole-dipole interactions and finite-size effects. The time after which $\overline{C_{l_{1}}(t)}$ is maximum, shown in Fig.~\ref{fig:fe_res3} (bottom), displays a minimum $t_{\overline{C},\mathrm{min}}\approx 0.317/\gamma_0$ at $k_0R\approx 0.60$ (to be compared to $0.438/\gamma_0$ for pure superradiance).

\begin{figure}[h!]
\begin{center}
\includegraphics[scale=0.6]{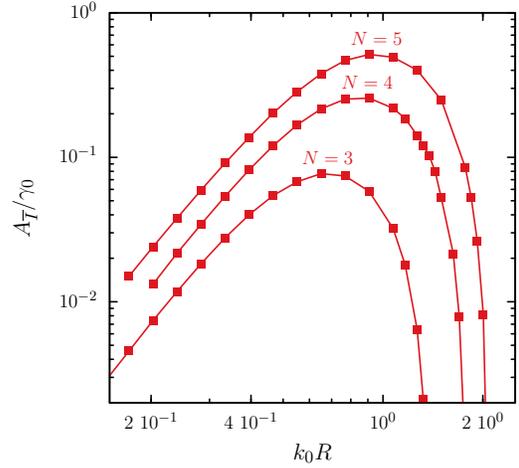}
\caption{Relative height $A_{\overline{I}}$ of the average radiated energy rate as a function of $k_0R$ for different number of atoms ($N = 3,4,5$ from bottom to top). All curves display a similar behavior and exhibit a maximum. The maximum of $A_{\overline{I}}$ and the optimal value of $k_0R$ both increase with the number $N$ of atoms : For $N = 3,4,5$, we have $A_{\overline{I},\mathrm{max}}\approx 0.078\,\gamma_0, \, 0.26\,\gamma_0, \, 0.52\,\gamma_0$ at $k_0R\approx 0.68,\, 0.85,\,0.94$, respectively.}
\label{fig:fe_res2}
\end{center}
\end{figure}

Our results on the intensity maximum for larger number of atoms ($N = 4,5$) are displayed in Fig.~\ref{fig:fe_res2}. A unique behavior emerges in which $A_{\overline{I}}$ attains a maximum for some value of $k_0R$. This shows that the mechanism leading to a maximum, i.e.\ the trade-off between dipole-dipole interactions and finite-size effects, is always present. Interestingly, the optimal value of $k_0R$ increases with the number of atoms : $k_0R\approx 0.68,\, 0.85,\,0.94$ for $N=3,4,5$.

Finally, we computed the average cooperativity parameter $\mathcal{C} = \rho \lambda^3/4\pi^2$ for the optimal values of $k_0R$ and obtained $\mathcal{C}\approx10^{-4}$. As this value is much smaller than $1$ but superradiance is nevertheless observed, $\mathcal{C}$ seems not suited to quantify cooperative effects when the number of atoms is very small.

\subsection{Decoherence-free state : subradiance}

\begin{figure}[h!]
\begin{center}
\includegraphics[scale=0.6]{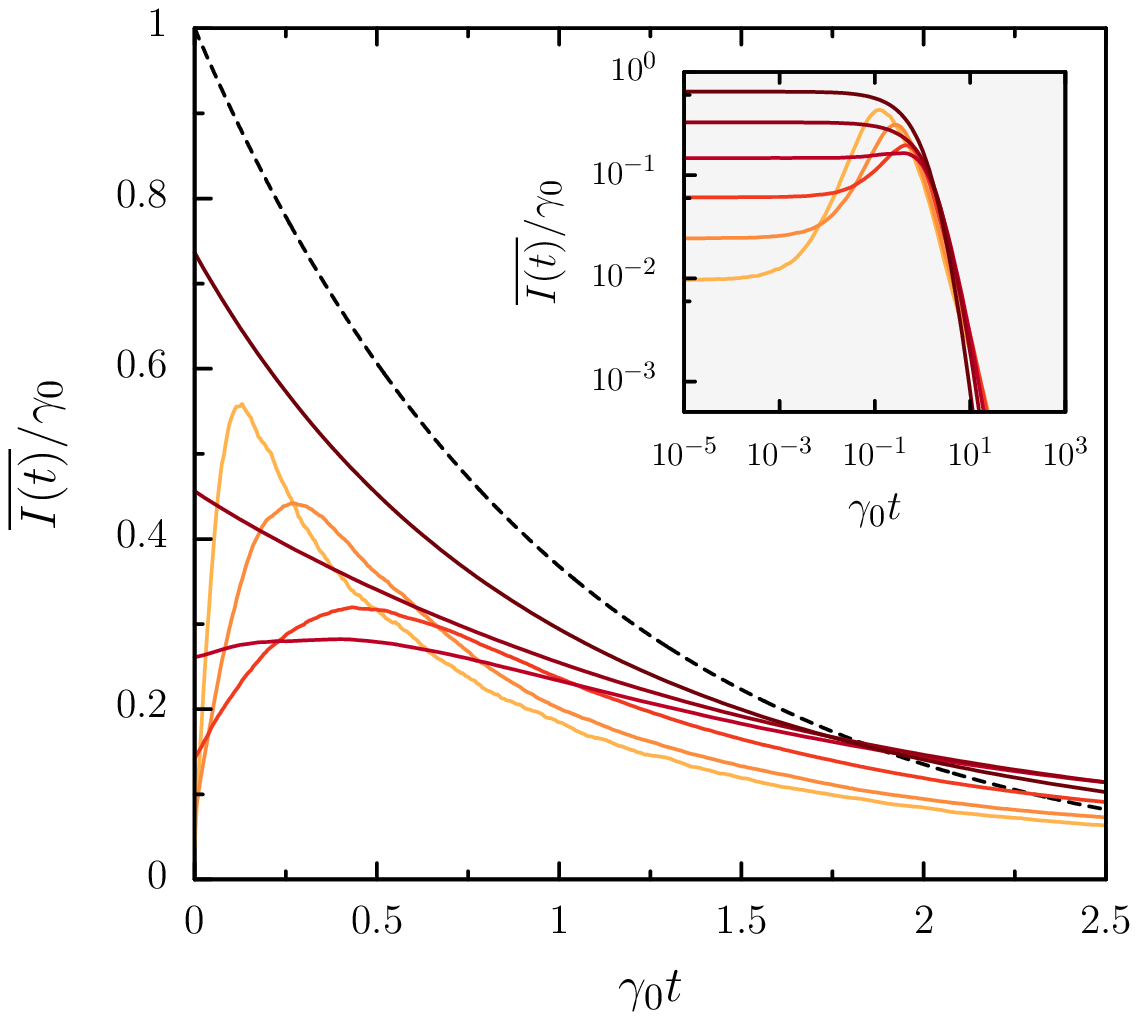}\\[6pt]
\includegraphics[scale=0.6]{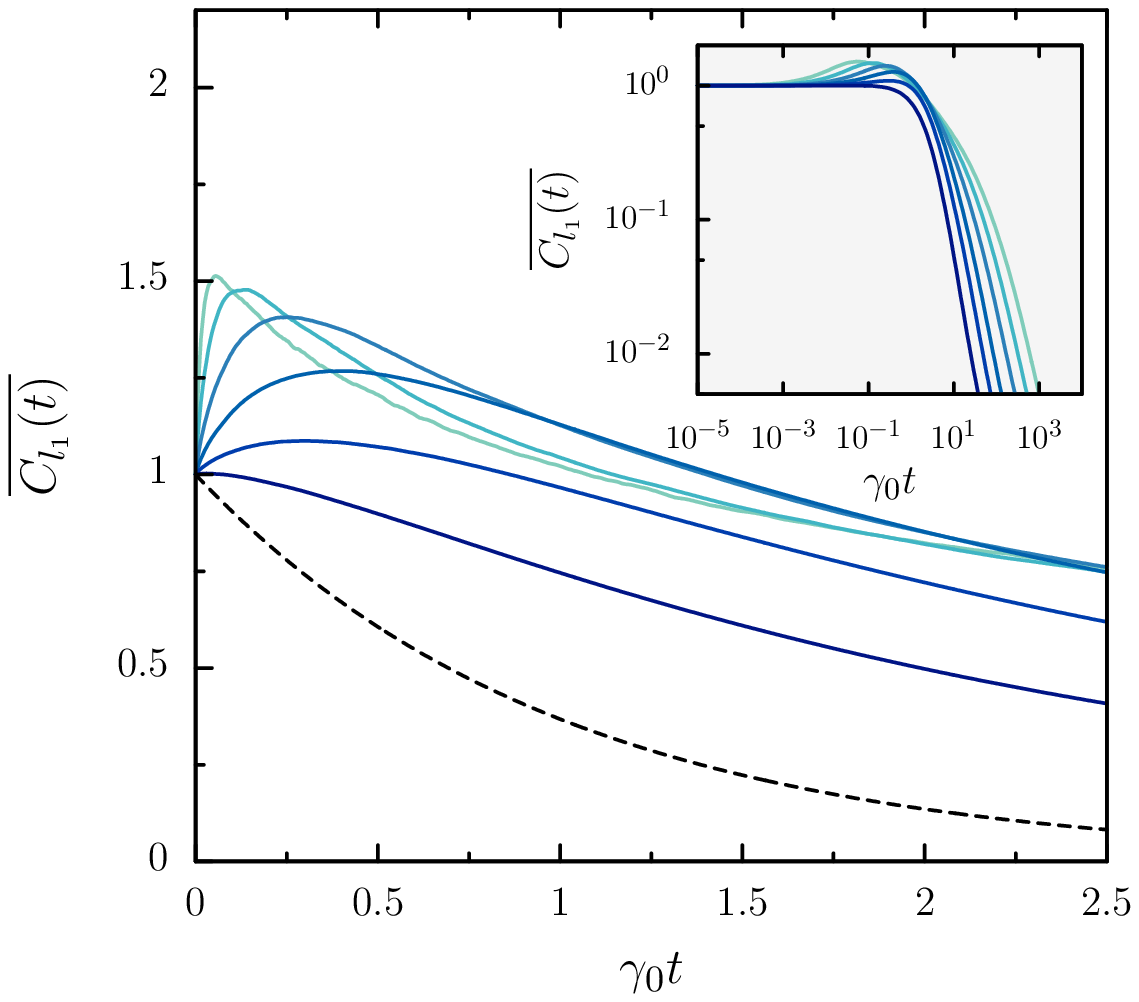}
\caption{Statistical average of the radiated energy rate ($\overline{I(t)}$, top) and the coherence ($\overline{C_{l_1}(t)}$, bottom) as a function of time for $3$ atoms initially in the subradiant state (\ref{eq:DFS2}). The average is taken over $10000$ random distributions for (from the lightest to the darkest) $k_0R =0.466,0.651,0.909,1.268,1.770$ and $2.470$.
The dashed black curve in both plots corresponds to the case of distant atoms ($k_0 R \to \infty$), for which $\overline{I(t)}$ (resp.\ $\overline{C_{l_1}(t)}$) decays exponentially as $\gamma_0 e^{-\gamma_0 t}$ (resp.\ $e^{-\gamma_0 t}$). For small $k_0R$, both the radiated energy rate and the coherence decrease algebraically with time ($\overline{I(t)} \propto t^{-1.2}$).}
\label{fig:DFS}
\end{center}
\end{figure}

\begin{figure}[h!]
\begin{center}
\includegraphics[scale=0.6]{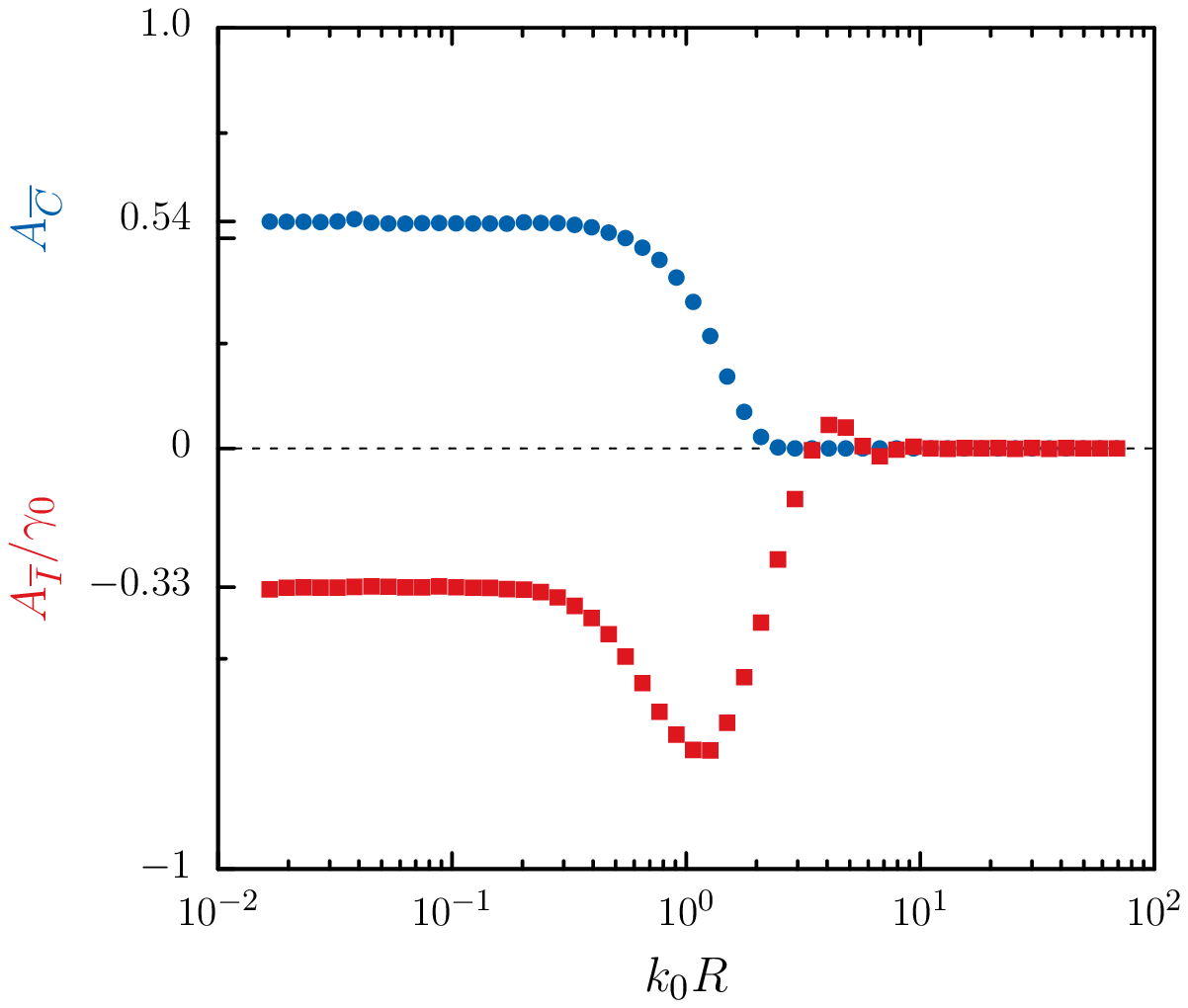}
\includegraphics[scale=0.63]{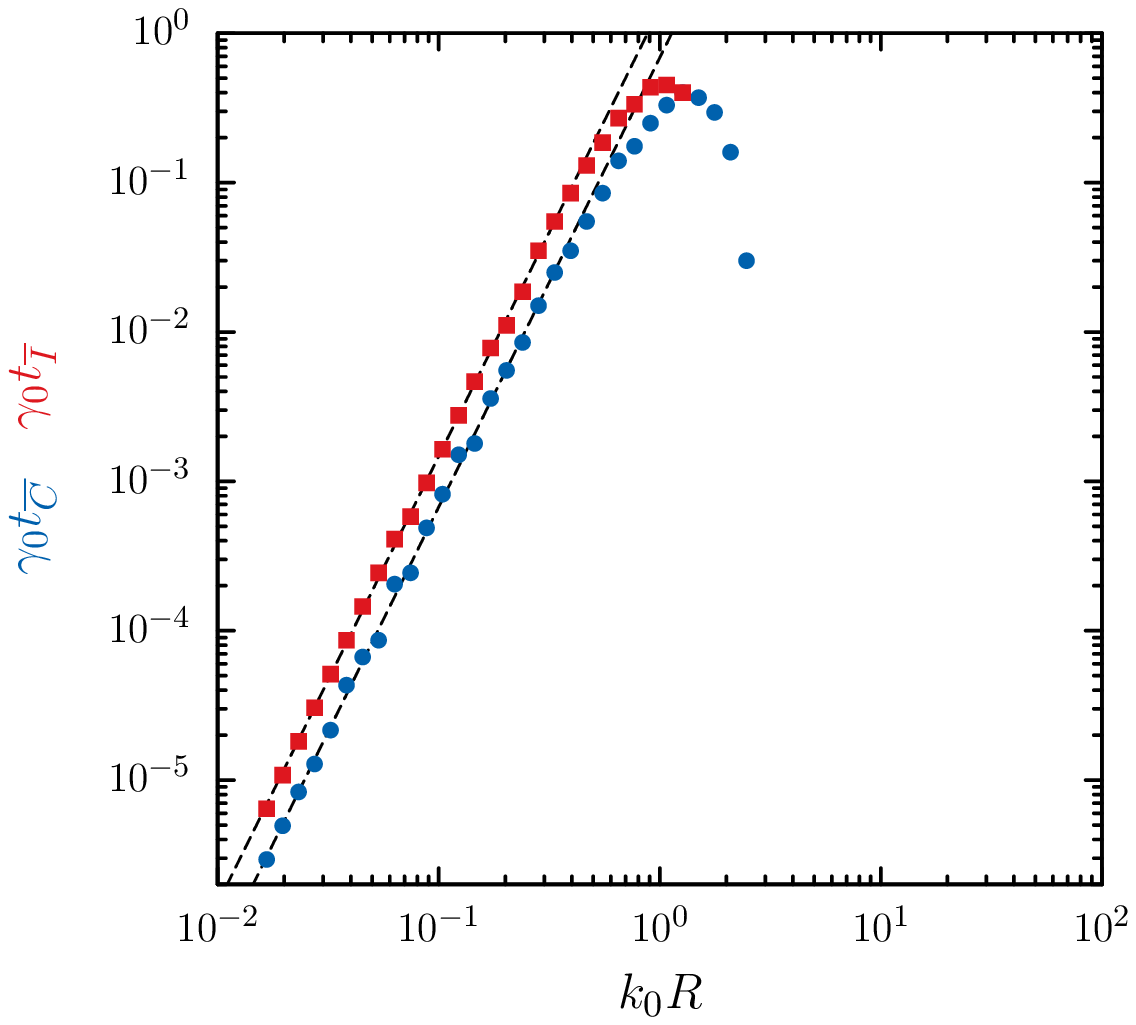}
\caption{Top: relative maxima of $\overline{I(t)}$ (squares) and $\overline{C_{l_1}(t)}$ (dots) as a function of $k_0R$.
Bottom: Power-law decrease of the times $t_{\overline{I}}$ (squares) and $t_{\overline{C}}$ (dots) at which $\overline{I(t)}$ and $\overline{C_{l_1}(t)}$ reach a maximal value, as a function of $k_0R$. The statistical average has been taken over $10000$ random atomic distributions. For sufficiently large values of $k_0R$, the pulses disappear completely (see Fig.~\ref{fig:DFS}), and $t_{\overline{I}} = 0$ and $t_{\overline{C}} = 0$.}
\label{fig:pl_th_intcoh}
\end{center}
\end{figure}

A decoherence-free state (DFS) $\rho_{\mathrm {DFS}}$ is a state whose time evolution is purely unitary, which enforces the condition
\begin{equation}
\mathcal{D}\left[\rho_{\mathrm {DFS}}(t) \right] = 0 \quad\forall\,t.
\label{eq:DFScondition}
\end{equation}
Atomic systems put in DFS do not radiate and are for this reason also called dark states. However, it has been shown in~\cite{Kar07} that for a set of two-level atoms governed by the master equation (\ref{eq:me})--(\ref{eq:gammaij}) no such states exist, even when the quantum fluctuations of the atomic positions are taken into account~\cite{Dam16,Dam16b}, except for the ground state or in the pure superradiant regime [see Eq.~(\ref{puresup})]. The absence of DFS means that the atoms will always release their internal energy in the environment to end up in the ground state $\ket{g_1\ldots g_N}$. In order to characterize the dynamics of this release, we focus on a system of three atoms initially in the state
\begin{equation}
\ket{\psi} = \frac{1}{\sqrt{2}} \left( \ket{g_1e_2g_3} - \ket{g_1g_2e_3} \right). 
\label{eq:DFS2}
\end{equation}
The state (\ref{eq:DFS2}) is separable with respect to the first atom and antisymmetric under exchange of the second and third atoms. In the pure superradiant regime ($\gamma_{ij} \approx \gamma_0$ and $f_{ij} \approx f_0 \;\; \forall i,j$), this state is decoherence-free. In the opposite limit of distant atoms ($\gamma_{ij} \approx \gamma_0 \delta_{ij}$ and $f_{ij} \approx 0 \;\; \forall i,j$), the averaged radiated energy rate (\ref{eq:int}) and coherence (\ref{eq:coh}) decay both exponentially as $\gamma_0 e^{-\gamma_0 t}$ and $e^{-\gamma_0 t}$, respectively. We computed them in the intermediate regime where finite size effects compete with dipole-dipole interactions. Our results are shown in Fig.~\ref{fig:DFS}. For $k_0 R\lesssim 1$, the radiated energy rate and coherence take the form of a pulse, and decrease algebraically at large times. In particular, the radiated energy rate decreases according to $\overline{I(t)} \propto t^{-1.2}$. When $k_0R$ increases, the pulse flattens and turns into the exponential decay reminiscent of independent spontaneous emissions. As previously, we define the superradiant pulse relative maximum $A_{\overline{I}}$ and the coherence pulse relative maximum $A_{\overline{C}}$ as in Eqs.~(\ref{Abar}) and (\ref{Cbar})  with $I_0 = \gamma_0$ and $ C_{l_1,0}=1$ (these are the initial values corresponding to independent spontaneous emission). A negative value for $A_{\overline{I}}$ means that the radiated energy rate is smaller than the single-atom spontaneous emission rate $\gamma_0$ at any times. Figure~\ref{fig:pl_th_intcoh} shows $A_{\overline{I}}$ and $A_{\overline{C}}$ (top) and $t_{\overline{I}}$ and $t_{\overline{C}}$ (bottom) as a function of $k_0R$. For small $k_0R$, the relative maxima are independent of $k_0R$  ($A_{\overline{I}} \approx -0.33\gamma_0$ and $A_{\overline{C}} \approx 0.54$). However, the smaller $k_0R$ is, the faster the maxima are reached. The times $t_{\overline{I}}$ and $t_{\overline{C}}$ decay both as $\propto (k_0R)^{-3} $ for small $k_0R$. This can be explained by the fact that, when $k_0R\lesssim 1$, the dynamics is dominated by dipole-dipole interactions which trap the excitations and lead to a reduction of radiated energy rate by a factor $\sim (k_0 R)^{3}$. 
Interestingly, $A_{\overline{I}}$ exhibits a minimum at $k_0R \approx 1$ and a maximum at $k_0R \approx 4$. The minimum (resp.\ maximum) indicates that the state is the most (resp.\ less) subradiant for this value of $k_0R$. In fact, in the region where $A_{\overline{I}}$ is positive, the state exhibits even a superradiant behaviour, i.e.\ an enhanced photon emission probability at short times.

\section{Conlusion}

In this paper, we studied the interplay between finite-size effects and dipole-dipole interactions on the cooperative dynamics of two-level atoms coupled to the electromagnetic field in vacuum. We first investigated the case of an initial fully-excited state which is known to lead to superradiance in the small sample limit (superradiant limit). Our statistical approach allowed us to observe a reduction of the radiated energy rate, as in previous works~\cite{Fri72, Fri74, Cof78, Gro82, Fre86, Fen13}, but also to show the existence of an optimal value of $k_0R$ for which superradiance is the less affected. This optimal value, of the order of $1$ for $N=3,4,5$ and increasing with $N$, was shown to result from the competition between finite-size effects and dipole-dipole interactions which is the most pronounced when the size of the atomic sample is of the order of the radiation wavelength. We also investigated the time evolution of the coherence of the atomic state and showed again the existence of a value of $k_0R$ (slightly different from the one for the energy rate) for which the coherence is maximum.
We then moved our focus to a three-atom system initially in a subradiant state which is dark in the superradiant limit. We found that the dynamics for different values of $k_0R$ displays common features:  for $k_0R\lesssim 1$, the energy rate and coherence reach a maximum and decrease algebraically at large times. The smaller $k_0R$ is, the faster the maximum is reached. For $k_0R \gtrsim 1$, exponential decays are recovered as is typical for independent spontaneous emissions. Surprisingly, for the range of intermediate values $3\lesssim k_0R \lesssim 5$, the subradiant state (\ref{eq:DFS2}) exhibits superradiance.

\ack FD would like to thank the F.R.S.-FNRS for financial support.
FD is a FRIA grant holder of the Fonds de la Recherche Scientifique-FNRS (F.R.S.-FNRS). Computational resources have been provided by the Consortium des \'Equipements de Calcul Intensif (C\'ECI), funded by the F.R.S.-FNRS under Grant No. 2.5020.11.

\appendix
\setcounter{section}{1}
\section*{Appendix : Methods}

In this section, we present further details about our computations and analytical calculations. We also discuss the averaging procedure used in this work to obtain statistical quantities such as $A_{\overline{I}}$ and $A_{\overline{C}}$.  

\subsection{Solving the master equation}

A common approach to solve a master equation like Eq.~(\ref{eq:me}) is to work in the dressed-states basis obtained from the diagonalization of the conservative part~\cite{Tor13, Fre86, Fre87, Fre04}. Some authors prefer to work instead in the basis formed by the eigenvectors of the non-unitary part, see e.g.~\cite{Bri93}. In both cases, the motivation is to split the global Hilbert space into orthogonal subspaces between which no coherences can be created during the time evolution. With this idea in mind, we choose in this work to gather states with the same number of excitations $n$ and write the $N$ two-level atom density matrix in the basis $\{\ket{e_1e_2\ldots e_{N-1}e_N},$ $\ket{e_1e_2\ldots e_{N-1}g_N},$ $\ket{e_1e_2\ldots g_{N-1}e_N},$ $\ldots,\ket{g_1g_2\ldots g_{N-1}g_N}\}$ as
\begin{equation}\label{rhoform1}
\rho(t) = \hspace*{-22pt}\bbordermatrix{
  ~n\to & N & N-1 & N-2 & \cdots & 0 \cr
  ~ & \square_1 & 0 &  \cdots &  & 0\cr
  ~ &  0 & \square_{C_N^{N-1}}  & 0 & \cdots &  \cr
  ~ &  \vdots & 0 & \square_{C_N^{N-2}} & \ddots  & \vdots\cr
  ~ &  & \vdots & \ddots & \ddots & 0 \cr
  ~ & 0 &  & \cdots & 0 &  \square_1\\}
\end{equation}
where $\square_k$ stands for a block of dimension $k$ and $C_{N}^{n} = N!/(N-n)!\, n!$. The form (\ref{rhoform1}) of the atomic density matrix is retained for all times $t$ because i) dipole-dipole interactions conserve the excitation number and couple elements within each block, and ii) the dissipative part only couples density matrix elements within different blocks and does not create coherences between blocks. Hence, all matrix elements outside the blocks, which are initially zero for the states considered in this work, remain zero at any time. The number of matrix elements among the $2^{2N}$ which need effectively to be accounted for in the case of an initial fully excited state is thus $\sum_{n = 0}^N C_{N}^{n}$ (which is equal to $6$ for $N=3$, $20$ for $N=4$ and $70$ for $N=5$). When the system is initially in the subradiant state (\ref{eq:DFS2}), it contains at most one excitation and $\rho(t)$ only involves $9$ (real) variables. 

To solve the master equation (\ref{eq:me}), it is first cast into a system of first-order coupled differential equations
\begin{equation}
\frac{d}{dt}\overrightarrow{\rho}(t) = A\, \overrightarrow{\rho}(t)
\label{eq:mev}
\end{equation}
where $A$ is square matrix and $\overrightarrow{\rho}(t)$ is the vectorization of the density matrix (\ref{rhoform1}). The system (\ref{eq:mev}) is then solved via $\overrightarrow{\rho}(t) = e^{At}\,\overrightarrow{\rho}(0)$ with $\overrightarrow{\rho}(0)$ the initial state. The radiated energy rate and coherence are subsequently computed using Eqs.~(\ref{eq:int}) and (\ref{eq:coh}). 

\subsection{Averaging procedure}

We generate random atomic distributions in spherical containers with prescribed average interatomic distance $R$. We do this by picking random positions within a sphere of arbitrary radius and rescaling all interatomic distances by their average. Then we multiply atomic positions by the prescribed $R$. Note that we discard distances smaller than the Bohr radius ($a_0$), since the divergence ($\propto r_{ij}^{-3}$) of the static dipole-dipole interaction between atoms is not physical when the atomic electron clouds overlap at distances $r_{ij} < a_0$~\cite{Bro08, Fen14}. The master equation (\ref{eq:me}) depends on the adimensional parameters $k_0 r_{ij}$ and $\alpha_{ij}$ (angle between the vector $\mathbf{r}_{ij}$ connecting atoms $i$ and $j$ and the dipole moment $\mathbf{d}_{eg}$) through the coefficients $\gamma_{ij}$ and $f_{ij}$ given in Eqs.~(\ref{eq:gammaij}) and (\ref{eq:fij}). In our simulations, $k_0$ is kept fixed and $\{\mathbf{r}_{ij}\}$ are varied.

\subsection{Single realizations of atomic distributions}

\begin{figure*}[h!]
\begin{center}
\includegraphics[scale=0.6]{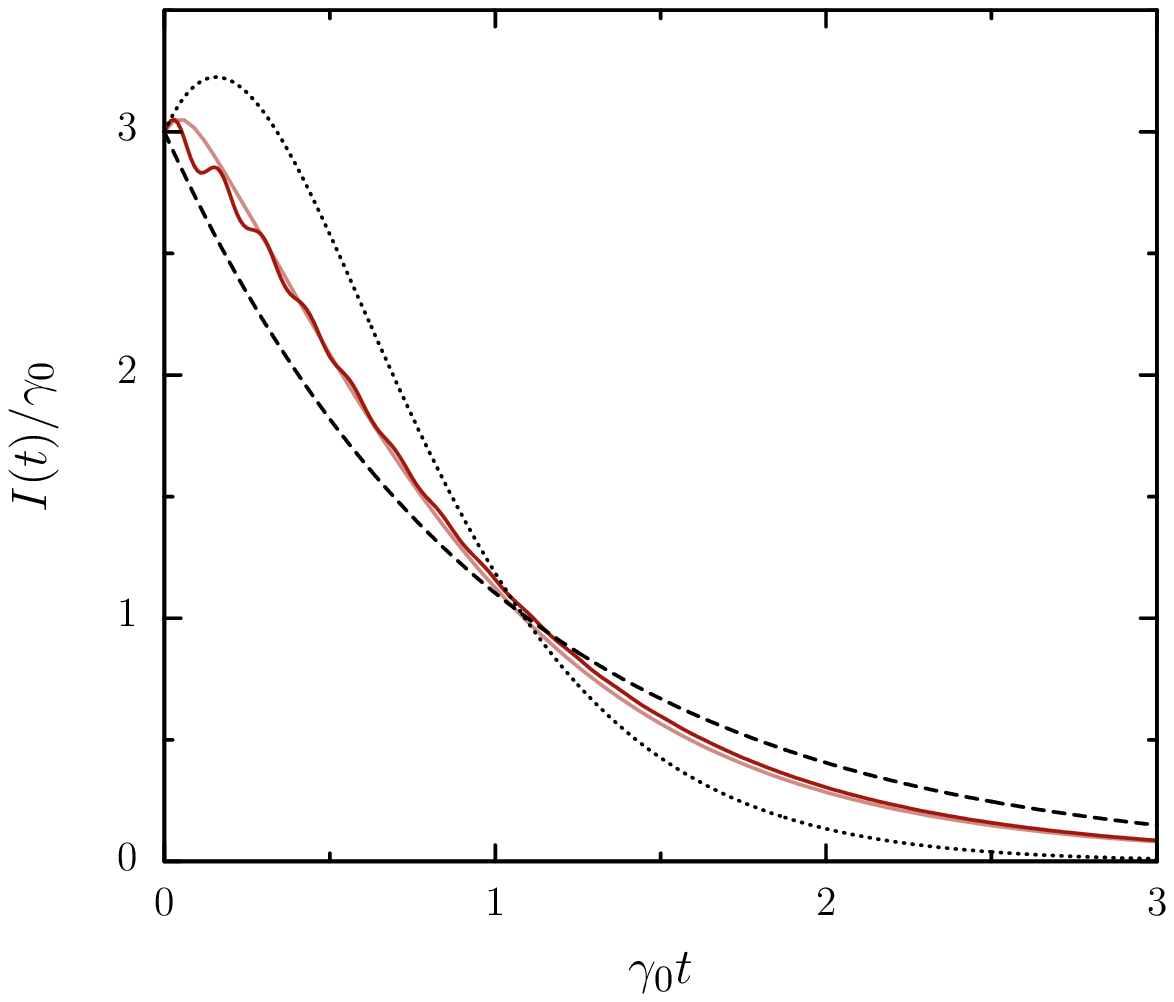}\qquad \includegraphics[scale=0.6]{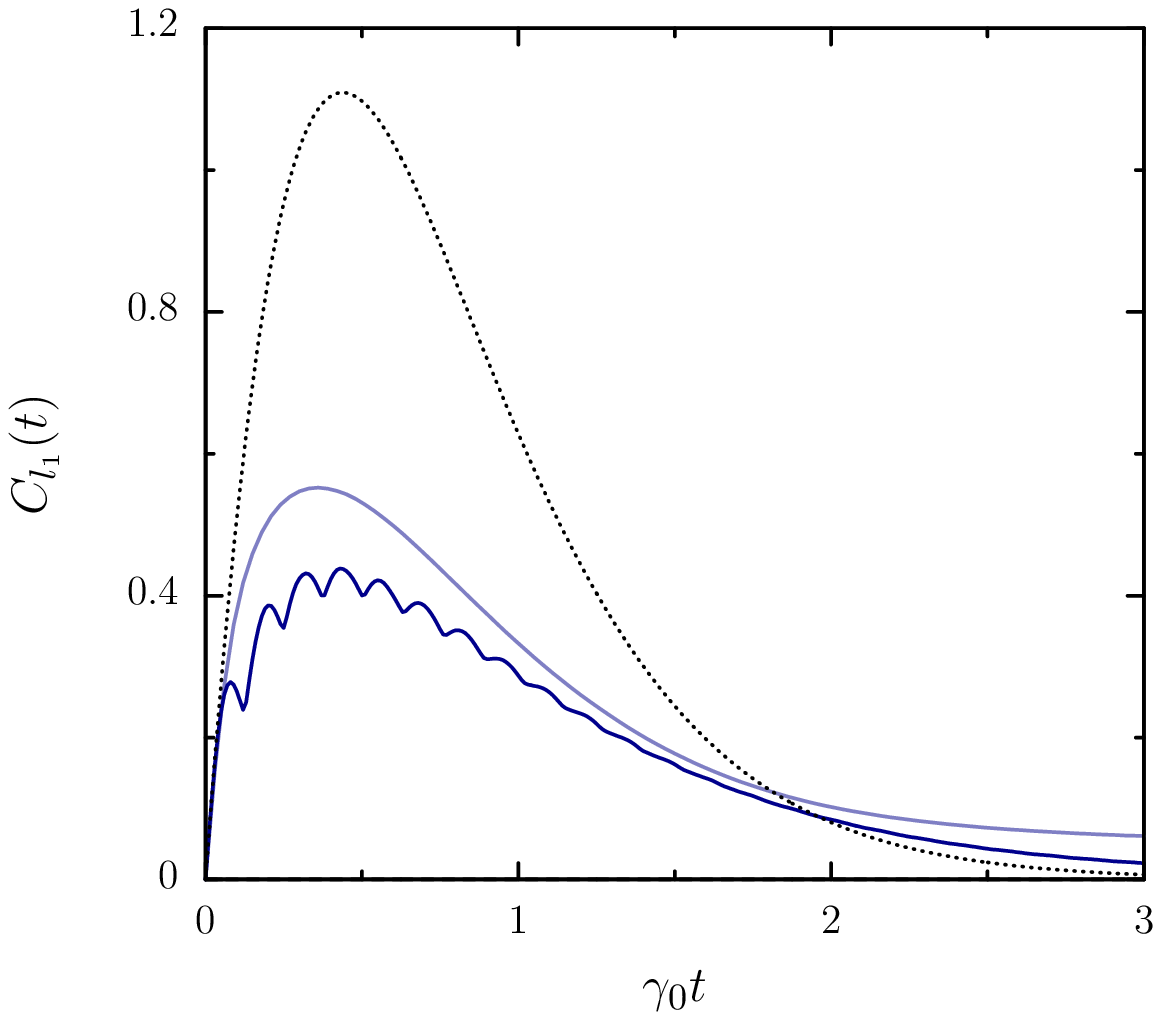}
\caption{Radiated energy rate $I(t)$ (left) and coherence $C_{l_1}(t)$ (right) obtained from the solution of the master equation without averaging over random atomic distributions ($N=3$, fully excited state). The dark red/blue oscillating curve corresponds to a single random realization of the atomic distribution with $k_0R \approx 0.466$. The light red/blue curve corresponds to the average values $\overline{I(t)}$ and $\overline{C_{l_{1}}(t)}$. The dotted and dashed curves show respectively the case of spatially close atoms with identical dipole-dipole shifts (superradiant regime) and distant atoms (independent spontaneous emission).}\label{fig:FE_nsim3}
\end{center}
\end{figure*}

\begin{figure*}[h!]
\begin{center}
\includegraphics[scale=0.6]{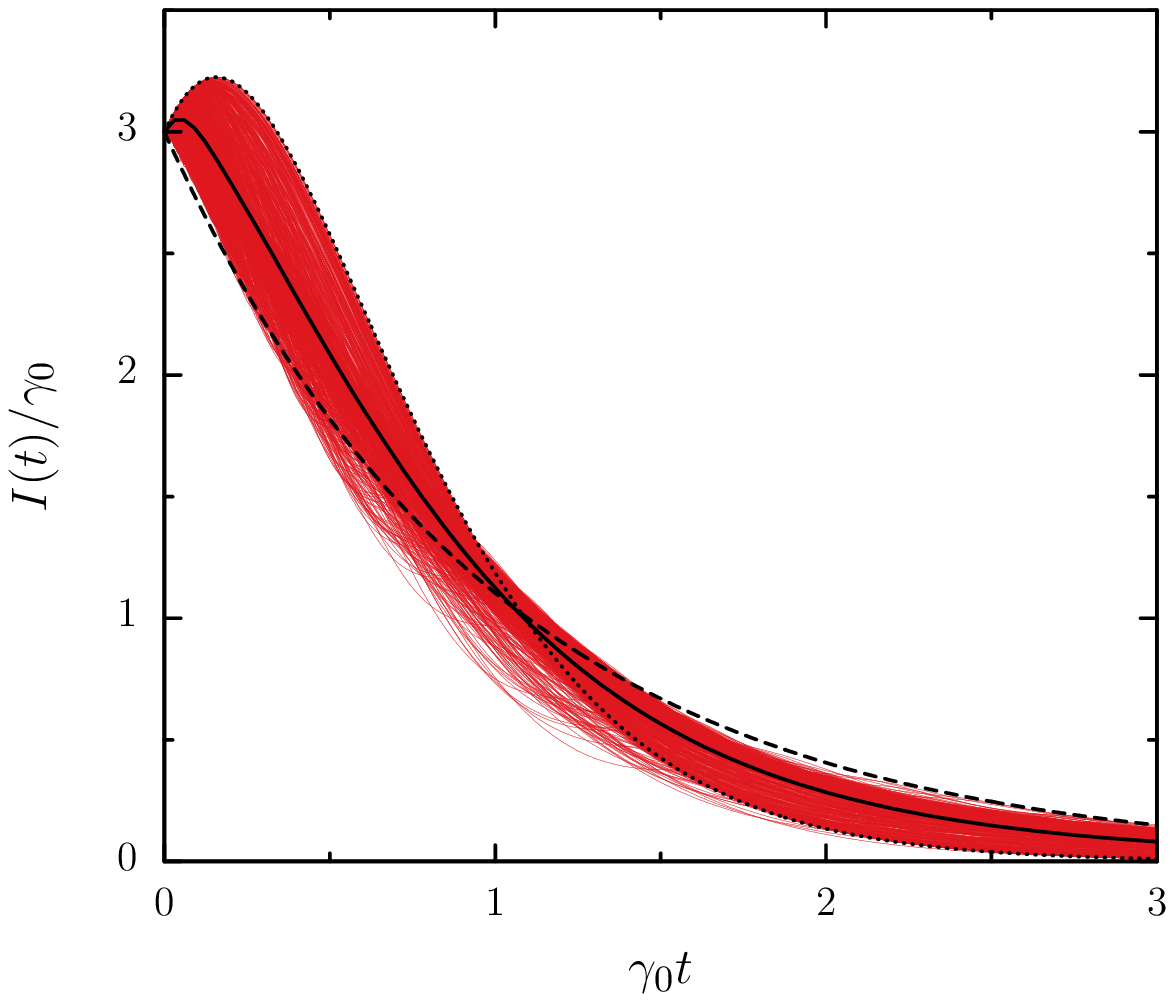}\qquad
\includegraphics[scale=0.6]{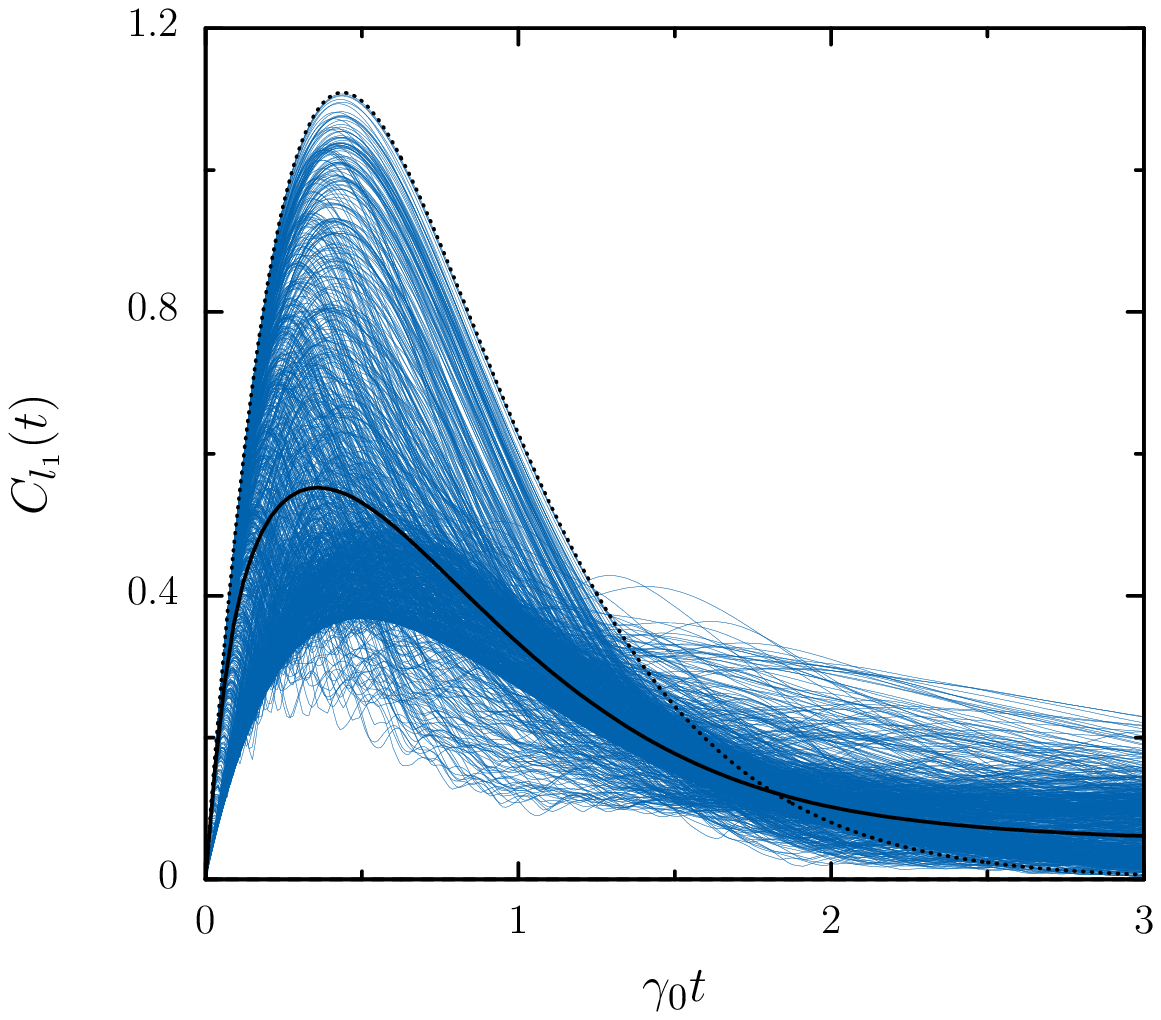} 	
\caption{Radiated energy rate $I(t)$ (left) and coherence $C_{l_1}(t)$ (right) for $1000$ different random atomic distributions ($N=3$, fully excited state) with $k_0R \approx 0.466$. The solid black curves correspond to the average values $\overline{I(t)}$ and $\overline{C_{l_{1}}(t)}$. The dotted and dashed curves show respectively the case of spatially close atoms with identical dipole-dipole shifts (superradiant regime) and distant atoms (independent spontaneous emission).}
\label{fig:FE_nsim1000}
\end{center}
\end{figure*}

From a single (random) realization of the atomic distribution (no average), the radiated energy rate and the coherence generally displays oscillations, as shown in Fig.~\ref{fig:FE_nsim3} with $k_0R \approx 0.466$. These oscillations, also known as beats, were first pointed out by Richter \cite{Ric90} and are due to unequal dipole-dipole shifts. When the dipole-dipole shifts are almost equal, the oscillations are less pronounced and $I(t)$ tends to the pure superradiant intensity. We also show in Fig.~\ref{fig:FE_nsim1000} a thousand curves of the radiated energy rate for random distributions and the coherence with $k_0R\approx 0.466$. In both cases, the curves corresponding to pure superradiance form the envelope of the distribution of curves. 

\end{document}